\documentclass[superscriptaddress,aps,pra,twocolumn,showpacs,longbibliography]{revtex4-2}
\usepackage[utf8]{inputenc}
\usepackage{graphicx,amsmath,amsfonts,amssymb}
\usepackage{color}
\usepackage[colorlinks=true, allcolors={blue}]{hyperref}
\usepackage{graphicx}
\usepackage{epstopdf}
\usepackage{float}
\usepackage{subfigure}

\newcommand{\orcid}[1]{\href{https://orcid.org/#1}{\includegraphics[width=7pt]{orcid.png}}}

\begin{document}

\title{Versatile probe state preparation via generalized measurements for quantum sensing and thermometry}

\author{Jonas F. G. Santos}
\email{jonassantos@ufgd.edu.br}
\affiliation{Faculdade de Ci\^{e}ncias Exatas e Tecnologia, Universidade Federal da Grande Dourados, Caixa Postal 364, CEP 79804-970, Dourados, MS, Brazil}

\author{Shanhe Su}
\email{sushanhe@xmu.edu.cn}
\affiliation{Department of Physics, Xiamen University, Xiamen 361005, People's Republic of China}

\author{Moises Rojas}
\email{moises.leyva@ufla.br}
\affiliation{Departamento de Física, Universidade Federal de Lavras, Caixa Postal 3037, 37200-900 Lavras-MG, Brazil}

\begin{abstract}
    We investigate a probe state preparation protocol based on two non-selective generalized quantum measurements to enhance parameter estimation in single-qubit systems. By fine-tuning the measurement strengths, we demonstrate the ability to design a broad class of probe states, initially prepared in a thermal state, which can be optimized for specific estimation tasks. We apply this framework to characterize the decay rate and the temperature of a generalized amplitude damping channel. Our results show that the preparation protocol significantly modulates the quantum Fisher information for both parameters. Furthermore, we derive a general analytical relationship between the quantum Fisher information, thermodynamic susceptibilities, and Hamiltonian variance, valid even in the transient regime. This connection highlights the role of energy fluctuations and kinetic response in determining metrological precision. Finally, we briefly discuss a quantum circuit for experimental implementation using nuclear magnetic resonance techniques.
\end{abstract}

\maketitle

\section{Introduction}

 Quantum metrology and quantum sensing have been made relevant progress in recent years in theoretical and experimental scenarios. Their robustness is demonstrated by the wide field of applications, for instance, in gravimetry \cite{Markus_Advac202, Wei_PhysRevR25, Joachim_arxiv25, Cassens_PhysRevX25}, quantum biology \cite{Das2024-lg, Aslam2023-pj}, interferometry \cite{APeters_2001, Biedermann_PRA15}, and thermometry \cite{franzao2024, xdnc-tc2y, PortoPRA2025, PhysRevLett.127.190402, Mehboudi2019-xi}. According to Ref. \cite{RevModPhys.89.035002}, the term \textit{quantum} is typically used in protocols employing a \textit{proper} quantum object with discrete energies as probe system, and/or the protocol is accomplished by the presence of quantum coherence \cite{Frazao2024-mz, Carlos_PRA2025} or entanglement \cite{Pezze_PRL09, Seth_PRL06} in order to boost the sensitivity beyond classical results. From a theoretical point of view, a pertinent direction is how the laws of thermodynamics and the irreversibility concept impose limitations on general quantum parameter estimation and sensing protocols in the quantum regime \cite{PhysRevLett.128.200501, PhysRevLett.133.263201}. Moreover, theoretical advances have been made by using quantum sensors for probing dark matter \cite{cwx5-2n1y}. Experimentally, progress have been reported in different platforms, such as trapped ions \cite{Delakouras2024-im}, Rydberg atoms \cite{Wang2026-mu}, superconducting systems \cite{Merlet2021}, and nuclear magnetic resonance (NRM) \cite{vs19-fwdz}.

The degree of accuracy in any parameter estimation process basically depends at least on three considerations: \textit{i}) the probe state preparation protocol, \textit{ii}) the interrogation time, which is the interaction time between the probe and the system or environment to be probed, and \textit{iii}) the set of measurements to be implemented in order to acquire information about a given parameter $\theta$. In classical estimation theory, the sensitivity is bounded from below by the Cramér-Rao inequality, which establishes that the standard deviation of a unknown parameter $\theta$ is $\text{Var}\left[{\theta}\right] \geq \left[N I_\theta\left(x\right)\right]^{-1/2}$, with $N$ the number of independent experiments trials and $I_\theta$ the so-called classical Fisher information (CFI), directly associated with the conditional probability of observing a random value $x$ \cite{Helstrom1969Quantum, Holevo2011-nq}. Physically, the CFI quantifies how sensitive is the probability distribution for a given measurement strategy or, more precisely, for a given set of positive operator-valued measurements (POVM) \cite{Seth_PRL06, Giovannetti2011}, leading to a CFI dependent on the set of POVMs. By optimizing over all possible POVMs to find the upper limit for the CFI, we find the well-known quantum Fisher information (QFI), $\mathcal{I}_\theta \equiv \text{sup}_{\text{POVM}}\lbrace I_\theta\left(x\right)\rbrace$. Analogously, the QFI gives rise to the quantum Cramér-Rao inequality, which establishes the minimal possible variance by employing a quantum probe system \cite{Liu2020-zo,Seth_PRL06, Giovannetti2011}.

Quantum metrology aims to consider different properties of quantum probe states to achieve the maximum QFI as possible. Intrinsic quantum features as coherence \cite{Frazao2024-mz, Carlos_PRA2025}, entanglement \cite{Pezze_PRL09, Seth_PRL06}, non-Hermiticity \cite{xdnc-tc2y, PhysRevLett.131.160801}, and contextuality \cite{Jae2024} have been employed to improve the QFI. Furthermore, the Mpemba effect has been also related to a boost in quantum thermometry protocols \cite{Chattopadhyay2026-cf}. In particular, the probe state preparation protocol is a suitable step where these quantum features can be manipulated in the probe system. In this scenario, generalized quantum measurements \cite{PhysRevE.96.022108} have shown to be an interesting mechanism in different branches of quantum information science and its application, as in designing information-based quantum machines \cite{PhysRevA.106.022436, Santos2023-bm}, for studying indefinite causal order \cite{PhysRevA.107.012423}, in quantum randomness certification \cite{PhysRevApplied.22.044041}, and for quantum state discrimination \cite{Reitzner2024-wt}. A central point of working with generalized quantum measurements is the possibility of varying their strength from very weak to completely projective ones \cite{Mancino2018-lx, Pan2020-my}, opening the window to manipulate quantum states to broad desired tasks.

In this letter, we investigate the use of two non-commutative generalized measurements playing the role of the probe state preparation protocol in a parameter estimation scheme. More concretely, the probe system is assumed to be initially in a thermal state, and by fine-tuning the generalized quantum measurement strengths, we are allowed to design different probe states, which may be useful to achieve a higher QFI for different parameters. After explaining in details the preparation protocol, we apply the method for characterizing parameters associated with two pertinent quantum channels, namely the amplitude damping (AD) and the generalized amplitude damping (GAD). We show that our preparation protocol is suitable for designing different probe states, each of them corresponding to the best QFI for each parameter estimation. We also obtain simple relations for the QFI in terms of different thermodynamic susceptibilities, even in the transient regime, evidencing the thermodynamic role in performing the parameter estimation procedure. Finally, a brief discussion about experimental realization using nuclear magnetic resonance is also introduced. 

The work is organized as follows. In Sec. \ref{Single-qubit parameter estimation and probe state preparation protocol} the basic mathematical aspects of parameter estimation and the quantum Fisher information are presented. We also detail the probe preparation protocol based on the two non-selective generalized measurements. Section \ref{Estimating parameters of a  Markovian environment} focuses on applying the probe state protocol to a relevant example, namely, the generalized amplitude damping. For this channel in particular, an experimental implementation using NMR is briefly discussed in Sec. \ref{Possible experimental implementation using NMR}. Finally, in Sec. \ref{Conclusions and remarks} we present the conclusion and final remarks. 

\section{Single-qubit parameter estimation and probe state preparation protocol}\label{Single-qubit parameter estimation and probe state preparation protocol}

\subsection*{Single-qubit parameter estimation}

Suppose that a physical channel is described by a set of $n$ unknown parameters $\vec{\lambda} = \left(\lambda_1,...,\lambda_n\right)$. For an arbitrary parameter $\lambda_i = \lambda$, the classical Fisher information (CFI) $I_\lambda$ is the amount of information that an observable random variable $X$ carries through a set of measurements outcomes $\lbrace x_i\rbrace$, with density probability $p_i\left(\lambda\right)$. Physically, the CFI quantifies how sensitive the probe density probability is for small variation of the unknown parameter $\lambda$. On the other hand, the so-called quantum Fisher information (QFI) is formally defined as \cite{Liu2020-zo}
\begin{equation}
    \mathcal{I}_\lambda = \text{Tr}\left[\rho_\lambda L_\lambda^2\right],
    \label{qFI01}
\end{equation}
where $L_\lambda$ is symmetric logarithmic derivative (SLD) Hermitian operator, determined by 
\begin{equation}
    \partial_\lambda \rho_\lambda = \frac{1}{2}\lbrace \rho_\lambda, L_\lambda\rbrace,
\end{equation}
with $\lbrace A,B\rbrace$ the anticommutator. Here, $\rho_\lambda$ the probe quantum state, which encodes information about the parameter $\lambda$, due to the interaction with the quantum channel during an interrogation time.

By diagonalizing the probe state as $\rho_\lambda = \sum_i \rho_i|\psi_i\rangle \langle\psi_i|$, Eq. (\ref{qFI01}) can be expressed as 
\begin{equation}
    \mathcal{I}_\lambda = \sum_{i}\frac{\left(\partial_\lambda \rho_{i}\right)^2}{\rho_i} + 2\sum_{i\neq j} \frac{\left(\rho_i - \rho_j \right)^2}{\rho_i + \rho_j}|\langle \psi_i|\partial_\lambda \psi_j\rangle|^2,
    \label{qFI02}
\end{equation}
with the sum over all $\rho_i \neq 0$ and $\rho_i + \rho_j \neq 0$. The first term in Eq. (\ref{qFI02}) corresponds to the classical Fisher information, whereas the second term is the quantum contribution, accounting for overlapping between different eigentates and their derivatives with respect to the parameter $\lambda$.

In particular, for a probe state represented by a qubit in the Bloch sphere decomposition, 
\begin{equation}
    \rho_\lambda = \frac{1}{2}\left(\mathbb{I} + \vec{r}_\lambda \cdot \vec{\sigma}\right),
    \label{Bloch}
\end{equation}
with $\vec{r}_\lambda = \left(r_x\left(\lambda \right),r_y\left( \lambda\right),r_z\left(\lambda \right) \right)^T$ the real Bloch vector and $\vec{\sigma} = \left(\sigma_x, \sigma_y,\sigma_z \right)$ the Pauli matrices, the quantum Fisher information for any mixed probe state can be expressed as \cite{Zhong2013FisherRepresentation}
\begin{equation}
    \mathcal{I}_\lambda = |\partial_\lambda \vec{r}_\lambda|^2 + \frac{\left(\vec{r}_\lambda \cdot \partial_\lambda \vec{r}_\lambda\right)^2}{1 - |\vec{r}_\lambda|^2}.
    \label{qFI03}
\end{equation}

Additionally, the main feature of the QFI is that it allows to obtain the achievable lower bound of the mean-square deviation $\left(\delta\lambda\right)^2$ of the unbiased estimator for a given parameter $\lambda$. This is done by the so-called quantum Cramér-Rao (QCR) bound \cite{Helstrom1969Quantum, Holevo2011-nq} 
\begin{equation}
    \left(\delta\lambda\right)^2 \geq \frac{1}{N \mathcal{I}_\lambda},
    \label{bound}
\end{equation}
with $N$ the number of repeated experiments.

The QCR bound in Eq. (\ref{bound}) imposes a fundamental limit on the precision in the estimation of a given parameter $\lambda$. In this sense, Eq. (\ref{qFI03}) is a suitable and powerful expression to quantum parameter estimation protocols, where the probe system, a qubit, interacts with a quantum channel with one or more unknown parameters during an interrogation time, and we aim to investigate the best possible precision for a such protocol.

\subsection*{Probe state preparation protocol}

We here now focus on explaining in detail the probe state preparation framework, which consists of applying two consecutive generalized quantum measurements to an initially qubit prepared in a thermal state. The qubit is described by the Hamiltonian $H = -\left(\hbar \omega\right/2)\sigma_z$, with $\omega$ the transition frequency and $\sigma_z$ standing for the Pauli matrix. It is initially prepared in a thermal state such that $\rho_{\text{th}} = \exp\left[-\beta H\right]/Z$, with $Z =\mathrm{Tr}\left[\exp\left(-\beta H\right)\right]$ the partition function. 

The probe state preparation protocol is composed of two generalized quantum measurements. They are described by positive operator-valued measurements (POVMs), where this approach includes phenomena not completely represented by projective measurements, such as detectors with nonunit efficiency weak measurements, and measurements outcomes with additional randomness. We choose the set POVMs for the first generalized measurements to be $\mathcal{M}_{1}=\sqrt{1-p}|0\rangle\langle0|+|1\rangle\langle1|$ and $\mathcal{M}_{2}=\sqrt{p}|1\rangle\langle0|$, such that $\sum_{i=1}^2\mathcal{M}_{i}^\dagger\mathcal{M}_{i} = \mathbb{I}$ and the state after this measurement is $\rho_{1}=\sum_{i=1}^{2}\mathcal{M}_{i}\rho_{\text{th}}\mathcal{M}_{i}^{\dagger}$. Here, $p$ is a parameter controlling the strength of the corresponding measurement, with $p=0$ and $p =1$ representing no action and a completely projective measurement on the probe state, respectively. In the sequence, the second generalized measurement is introduced with the set of POVMs, $\mathcal{N}_{1} =|0\rangle\langle0|+\sqrt{1-q}|1\rangle\langle1|$ and $\mathcal{N}_{2}=\sqrt{q}|0\rangle\langle1|$, with $q$ controlling the strength of the second measurement and $\sum_{i=1}^2\mathcal{N}_{i}^\dagger\mathcal{N}_{i}  = \mathbb{I}$. The probe state after the second measurement is then $\rho_{\text{probe}} = \sum_{i=1}^{2}\mathcal{N}_{i}\rho_{1}\mathcal{N}_{i}^{\dagger}$. The explicit expression for the probe state after the preparation protocol is
\begin{equation}
    \rho_{\text{probe}}=\frac{1}{2}\left\{ \mathbb{I}+\mathcal{R}_{z}\left(p,q\right)\sigma_{z}\right\}, 
    \label{probe01}
\end{equation}
where $\mathcal{R}_{z}\left(p,q\right) \equiv q\left(1+p\right)-p+\left(1-p\right)\left(1-q\right)\tanh y_0$, with $y_0 \equiv \beta \hbar \omega/2 = \hbar \omega/\left(2k_B T_0\right)$ carrying the preparation protocol information, and $T_0$ the initial temperature of the probe state.

\begin{figure}[!h]
\includegraphics[scale=0.25]{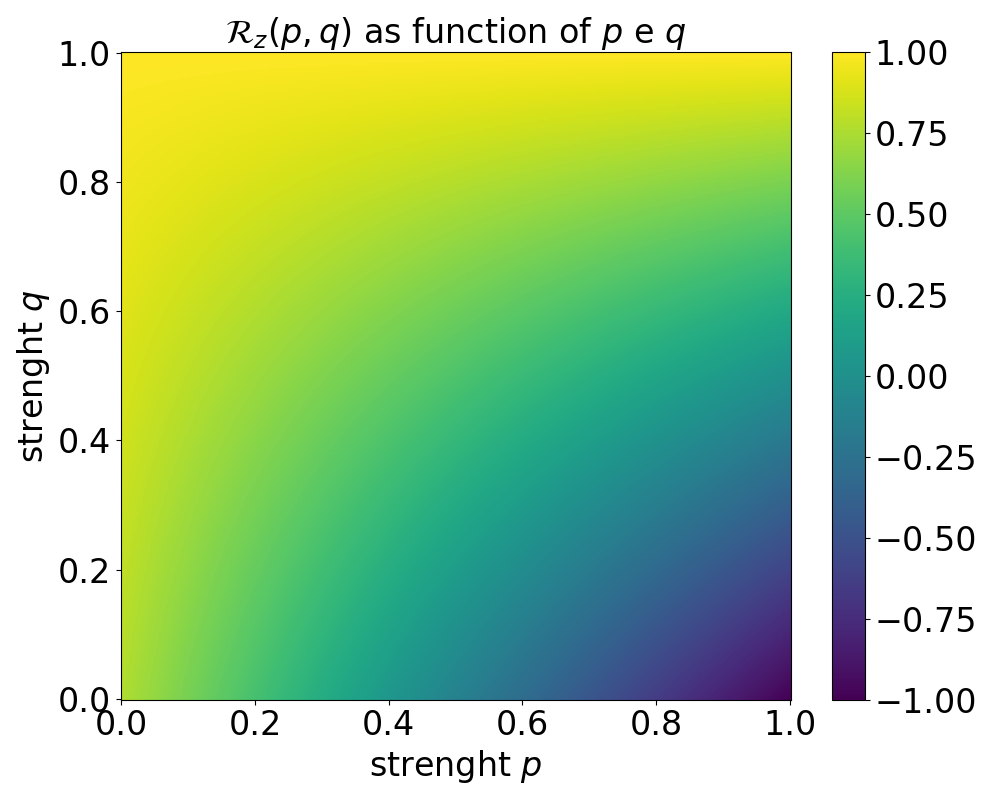}  
\caption{The contour plot of $\mathcal{R}_{z}\left(p,q\right)$ as functions of $p$ and $q$, corresponding to different probe state preparation. We have set $y_0 = 1$. While $\left(0,0\right)$ sets the initial thermal state, $\left(1,0\right)$ and $\left(0,1\right)$ define the probe state to be ground and excited state, respectively.}
\label{R_z}
\end{figure}

Figure \ref{R_z} illustrates $\mathcal{R}_{z}\left(p,q\right)$ as functions of $p$ and $q$, where each pair $\left(p,q\right)$ corresponds to a probe state. In particular, $\left(0,0\right)$, $\left(1,0\right)$, and $\left(0,1\right)$ set the probe state to be the initial thermal state, ground state, and excited state, respectively. Note that the current preparation protocol enables to cover a broad class of probe states, each of which may be suitable for different estimation processes.

\section{Estimating parameters of a  Markovian environment} \label{Estimating parameters of a  Markovian environment}

In order to illustrate how the proposed probe state preparation protocol works, we consider the generalized amplitude damping (GAD) channel. It is a paradigmatic single-qubit channel, using for describing noisy and thermal environments. The GAD channel is specified by two parameters, such that in this case $\vec{\lambda} = \left(\Gamma,T\right)$ with $\Gamma$ and $T$ the decay rate and the temperature, respectively. Both parameters are assumed to be fixed before, during, and after the interrogation time. Following the Born-Markov approximations \cite{breuer2002theory}, the master equation representing the probe state evolution during the interrogation time is given by
\begin{eqnarray}
    \frac{d \rho_{\text{probe}}}{dt} &=& -\frac{i}{\hbar}\left[H, \rho_{\text{probe}}\right]\nonumber\\
    &+& \Gamma\left(\bar{n}+1\right) \left(\sigma_-\rho_{\text{probe}} \sigma_+ - \frac{1}{2}\lbrace \sigma_+\sigma_-,\rho_{\text{probe}} \rbrace \right)\nonumber\\
    &+& \Gamma \bar{n}\left(\sigma_+\rho_{\text{probe}} \sigma_- - \frac{1}{2}\lbrace \sigma_-\sigma_+,\rho_{\text{probe}} \rbrace \right),
\end{eqnarray}
with $\sigma_+$ and $\sigma_-$ standing for the to raising and lowering operators, respectively, and $\bar{n}$ the respective average thermal number. The amplitude damping (AD) channel is recovered in the limit of $\bar{n} \rightarrow 0$.

Whenever a qubit is described using the Bloch sphere decomposition in Eq. (\ref{Bloch}), all information about the two-parameters family is encoded in the real Bloch vector $\vec{r} = \vec{r}\left(\Gamma,T,t\right)$ where $t$ corresponds to the interrogation time. From Ref. \cite{breuer2002theory, Prathik_Cherian2019-en} and using the probe state in Eq. (\ref{probe01}), we immediately identify  $r_x \left(\Gamma,T,t\right) =  r_y\left(\Gamma,T,t\right) = 0$, while 
\begin{equation}
    r_z\left(t\right) = \tanh\left(y_{eq}\right) - \bar{\Delta}\left(p,q\right) e^{-\Gamma t},
    \label{rzz}
\end{equation}
with $\bar{\Delta}\left(p,q\right) \equiv \tanh\left(y_{eq}\right) - \mathcal{R}_z\left(p,q\right)$ and $y_{eq} = \hbar\omega/\left(2k_B T\right)$. Note that the probe preparation protocol naturally affects the state dynamics during the interrogation time.

Assuming that $y_{eq}$ is independent of $\Gamma$ and taking the partial derivative of Eq. (\ref{rzz}) with respect to $\Gamma$, we obtain 
\begin{equation}
    \mathcal{I}_{\Gamma} = \frac{t^2 \overline{\Delta}(p,q)^2 e^{-2\Gamma t}}{1 - [\tanh(y_{eq}) - \overline{\Delta}(p,q)e^{-\Gamma t}]^2}.
\end{equation}
On the other hand, taking the partial derivative of Eq. (\ref{rzz}) with respect to $T$, we get 
\begin{equation}
    \mathcal{I}_T = \frac{\bar{\omega}^2\left(1 - e^{-\Gamma t}\right)^2 \text{sech}^2\left(\bar{\omega}/T\right)}{1 - \left[e^{-\Gamma t}\mathcal{R}_z\left(p,q\right) + \left(1 - e^{-\Gamma t}\right)\tanh\left(\bar{\omega}/T\right) \right]^2} \frac{1}{T^4}.
    \label{QFI_GAD}
\end{equation}

\begin{figure}[!h]
\includegraphics[scale=0.25]{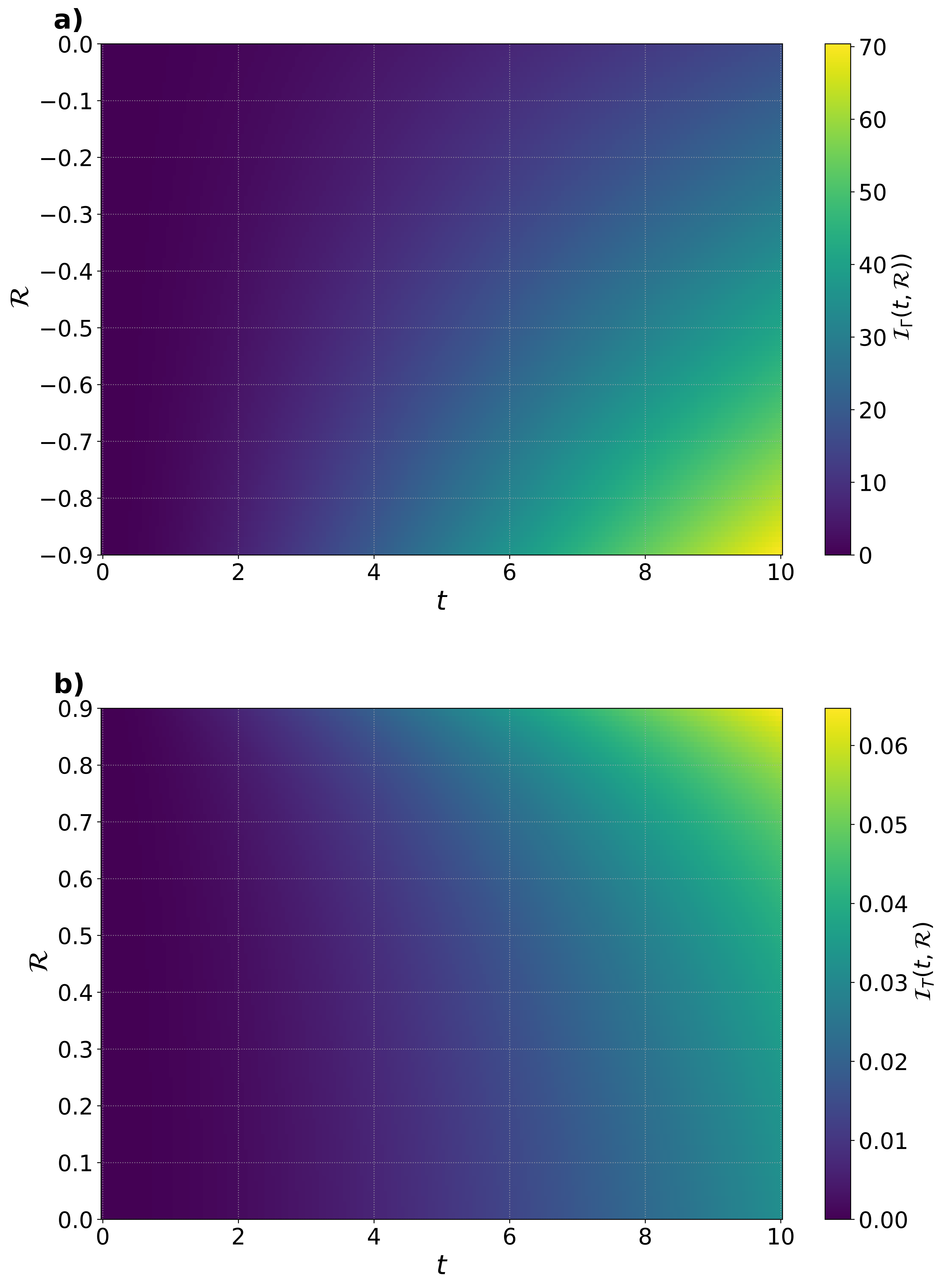}  
\caption{Quantum Fisher information. a) QFI for the estimation of $\Gamma$. Here, we assumed $y_{eq} = 1$ in order to see explicitly the effect of the probe state preparation. b) QFI for the estimation of $T$. Here, we defined $y_{eq} = 0.5$. Both QFIs are function of the interrogation time $t$ and for different probe state preparations defined by $\mathcal{R}$.  We have set $\hbar\omega = 1.0$ and $\Gamma = 0.05$.}
\label{QFIFigures}
\end{figure}

Figure \ref{QFIFigures} illustrates the behavior of $\mathcal{I}_\Gamma$ and $\mathcal{I}_T$ as functions of the interrogation time $t$ and the function representing different probe state preparations, $\mathcal{R}$. As pointed previously, the probe state preparation allows to design different states, each of them suitable for different parameter to be estimated, as illustrated in Fig. \ref{QFIFigures}, in which $\mathcal{R} \approx -1$ and $\mathcal{R} \approx 1$ provide higher QFI for $\Gamma$ and $T$, respectively. It should be empathized that in the present case, the best probe state depends also on the relation between the temperatures of the initial probe state and of the thermal environment to be interrogated.

Additionally to the QFI, the energy change is another relevant figure of merit in metrology, making the direct link with quantum thermodynamics. For the GAD channel, the energy change for an interrogation time $t$ reads
\begin{equation}
    \Delta U_{GAD} = -\frac{\hbar\omega }{2}\bar{\Delta}\left(p,q\right)\left(1 - e^{-\Gamma t}\right).
    \label{Energy}
\end{equation}

Since the energy change also depends of the two-parameters family, we can also introduce the thermodynamic susceptibility for an interrogation time $t$ as 
\begin{equation}
    \chi_{\theta_i}\left(t\right) = \frac{\partial \Delta U_{GAD}\left(t\right)}{\partial \theta_i},
    \label{suscep}
\end{equation}
with $\vec{\theta} = \left(\Gamma, T\right)$ the set of parameters to be estimated in both AD or GAD channels. 

Physically, while the $\chi_{T}\left(t\right)$ is the transient heat capacity and goes to the equilibrium heat capacity for an interrogation time $t\rightarrow\infty$, $\chi_{\Gamma}\left(t\right)$ represents the kinetic part of the susceptibility. Explicitly, the expressions are given by
\begin{eqnarray}
    \chi_{T}\left(t\right) &=& -\frac{\hbar \omega}{2}\left(1- e^{-\Gamma t}\right)\partial_T \tanh\left(y_{eq}\right) \nonumber\\
    \chi_{\Gamma}\left(t\right) &=& -\frac{\hbar \omega}{2}\bar{\Delta}\left(p,q\right)\partial_\Gamma\left(1- e^{-\Gamma t}\right),
\end{eqnarray}
and we note that $\chi_{\Gamma}\left(t\right)$ is sensitive to the probe state preparation protocols, whereas $\chi_{T}\left(t\right)$ is not and quantifies basically the system ability to gain or loss energy. 

Using Eq. (\ref{suscep}), we can derive an expression linking the QFI to the kinetic and thermodynamic susceptibilities and the variance of the Hamiltonian. It is given by 
\begin{equation}
    \mathcal{I}_{\theta_i} = \frac{\chi_{\theta_i}^2\left(t\right)}{\Delta H^2\left(t\right)},
    \label{QFI03}
\end{equation}
where $\Delta H^2(t) = \langle H^2 \rangle - \langle H\rangle^2 =\hbar^2\omega^2\left(1 - r_z\left(t\right)^2 \right)/4$. Here, we make the following observation: An analog expression to Eq. (\ref{QFI03}) for the temperature $T$ has been derived in the asymptotic limit of interrogation times $t \rightarrow\infty$, i.e., thermal equilibrium between the probe and the environment, for instance in Ref.\cite{Correa2015}. In the present case, Eq. (\ref{QFI03}) is the generalization for both $T$ and $\Gamma$ as well as for a transient interrogation time. Moreover, Eq. (\ref{QFI03}) together with Eq. (\ref{bound}) represents the temperature-energy uncertainty relation for nonequilibrium thermometry \cite{Miller2018-ga, PhysRevA.110.012211}, in the sense that the greater the energy fluctuation, the worse the estimate of the thermometric parameter. In addition, Eq. (\ref{QFI03}) shows explicitly that by fine-tuning the external controllable parameters, it may be useful to obtain a higher QFI even for short interrogation times.

\begin{figure}[!h]
\includegraphics[scale=0.4]{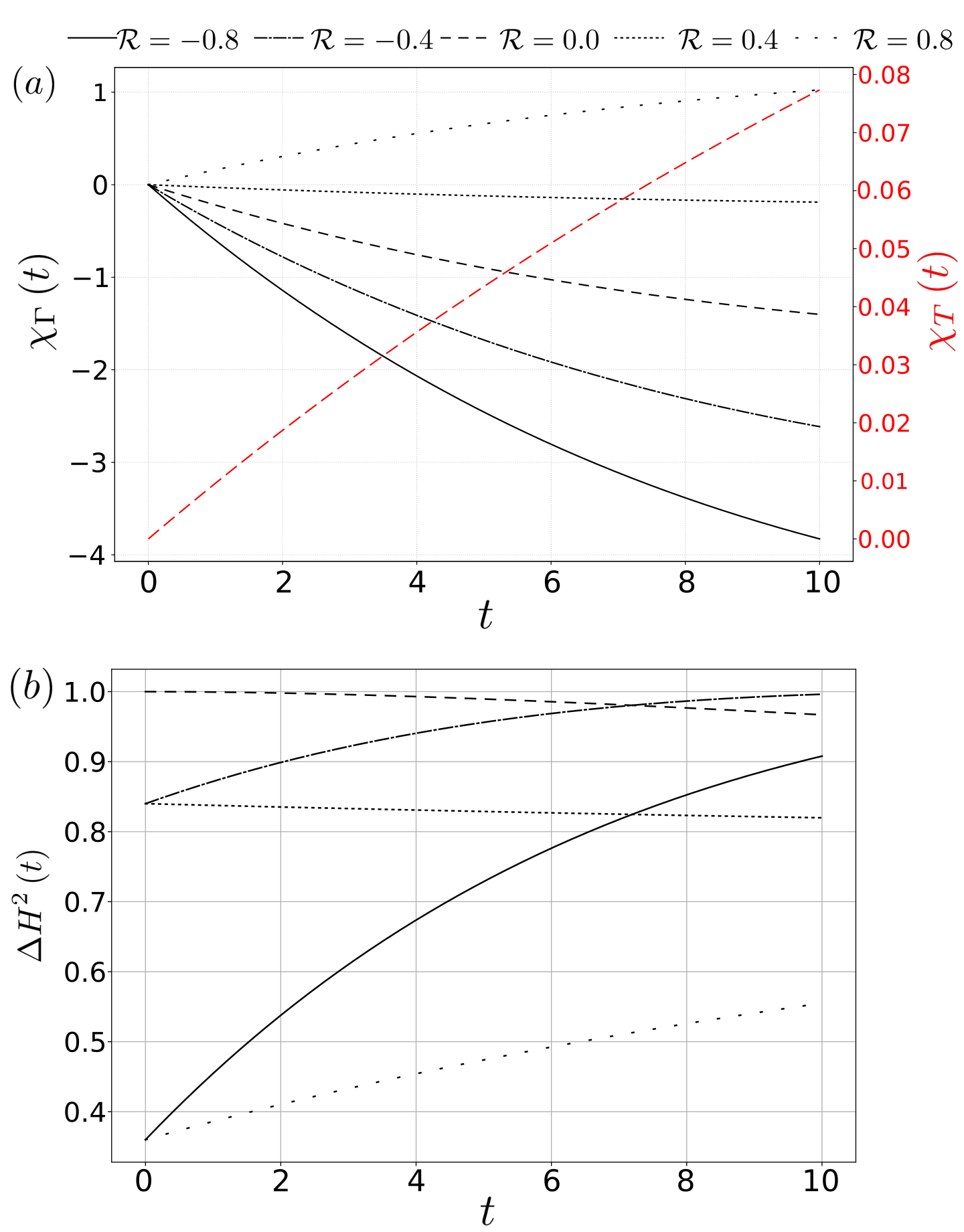}  
\caption{a) Kinetic and thermodynamic susceptibility as function of the interrogation time for different probe state preparations. b) Standard deviation as function of the interrogation time for different probe state preparations. We have set $\hbar\omega = 1.0$, $T = 1$ (in units of $\hbar\omega/k_B$), $\Gamma = 0.05$ and $y_0 = 1$.}
\label{Susc}
\end{figure}

In order observe how the probe state preparation protocol affect the kinetic and thermodynamic susceptibilities, Fig. \ref{Susc}-a)  illustrates $\chi_{\Gamma}\left(t\right)$ (solid lines) and $\chi_{T}\left(t\right)$ (dashed line) as functions of the interrogation time. On the other hand, Fig. \ref{Susc}-b) depicts $\Delta H\left(t\right)$ also as a function of the interrogation time. Both are for different values of $\mathcal{R}$ as indicated in the legend. In particular, for thermometry, despite $\chi_{T}\left(t\right)$ does not depend on $\mathcal{R}$, $\Delta H\left(t\right)$ does, which also show the role played by the probe state preparation in the temperature estimation.

While we focused on the GAD channel, the versatility of our probe preparation protocol suggests its immediate applicability to other relevant noise models, such as the class of Pauli channels \cite{PhysRevA.108.042612}, where the fine-tuning of $\mathcal{R}$ could similarly optimize the estimation of error probabilities in quantum computing architectures.

\section{Possible experimental implementation using NMR} \label{Possible experimental implementation using NMR}

Recently, the two generalized quantum measurements designed here as the preparation protocol for the probe state have been experimentally implemented in a quantum thermal machine model using NMR techniques \cite{PhysRevA.106.022436}. Based on this implementation, we here propose a four-qubit quantum circuit in order to experimentally test our quantum estimation protocol. In this scenario, the first qubit plays the role of the probe, while the second and third qubits work as ancillas to implement the first and the second generalized measurements, respectively. To complete, a fourth qubit acts as an ancilla to implement the channels to be probed.

\begin{figure}[!h]
\includegraphics[scale=0.02]{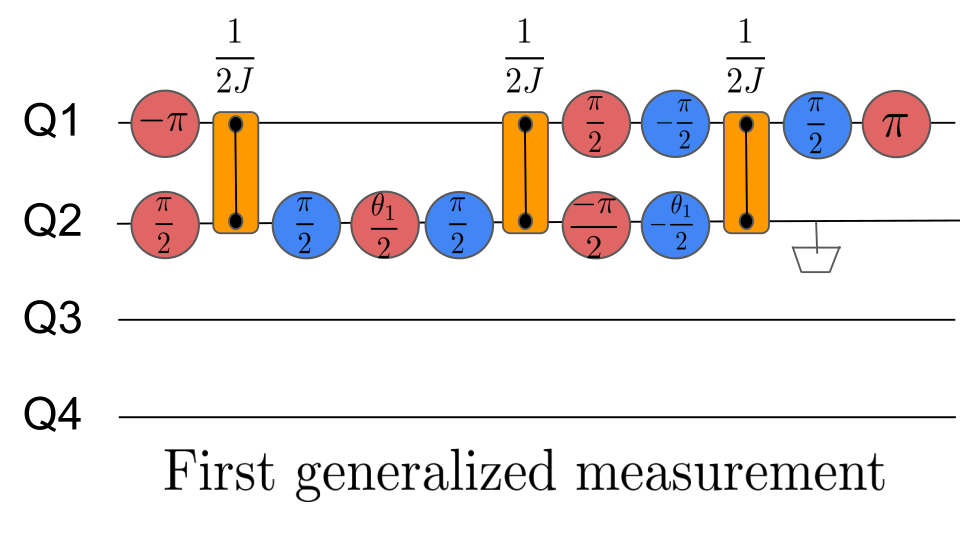}
\includegraphics[scale=0.02]{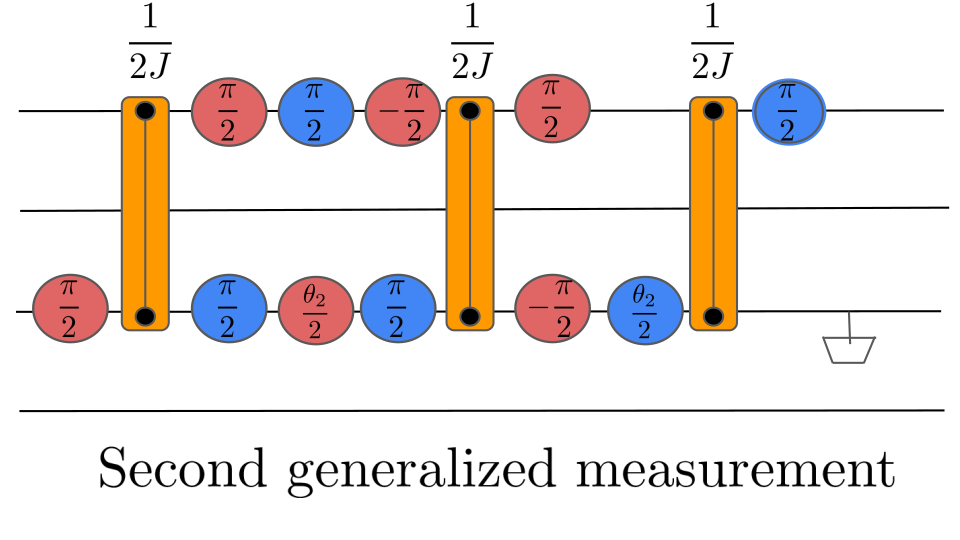}
\includegraphics[scale=0.02]{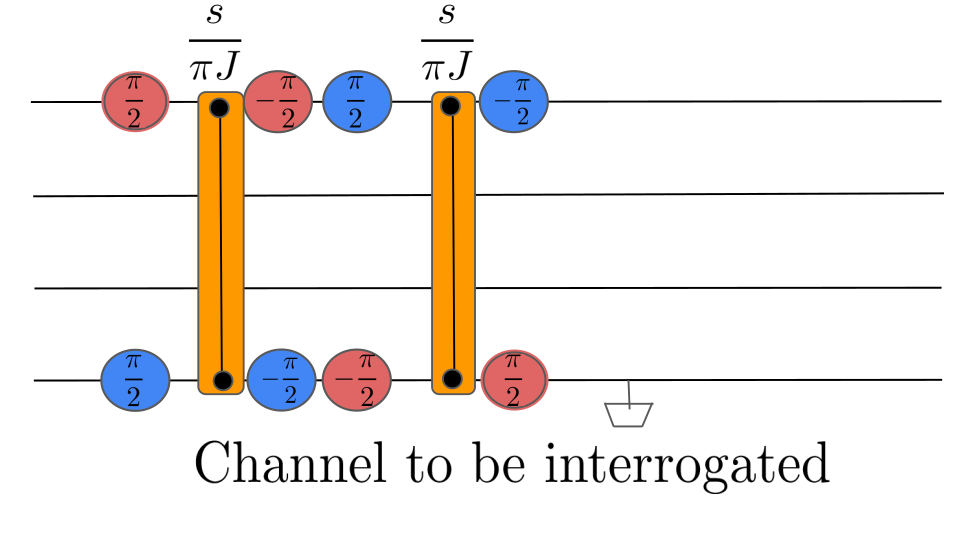}
\caption{Propose of pulses sequences to implement probe state preparation protocol and the AD or GAD channels. For the probe state preparation, we have two nonselective generalized measurements channels, $\mathcal{M}$ and $\mathcal{N}$. For $\mathcal{M}$ and $\mathcal{N}$, the measurement strength parameters $p$ and $q$ are tuned by the angles $\theta_1$ and $\theta_2$, respectively, with both angles ranging in $[0,\pi/2)$, according to $\theta_1 = \arccos\left(1 - 2p \right)$ and $\theta_2 = \arccos\left(1 - 2q \right)$. The qubits Q2, Q3, and Q4 are not observed, resulting in the desired nonselective operations channels (CPTP maps) acting on the Q1 (probe system). The parameter $s$ is related to the interrogation time $t$ by $\cos\left(s\right) = \exp\left[-\Gamma t/2\right]$. This setup for the quantum circuit is based on Ref. \cite{PhysRevA.106.022436}.}
\label{circuit_figure}
\end{figure}

A possible quantum circuit is illustrated in Fig. \ref{circuit_figure}, where the qubits are denoted by Q1 (probe), Q2 and Q3 (first and second generalized measurements), and Q4 (environment to be interrogated). The fourth qubit (Q4) serves as a controllable quantum simulator of the thermal environment, where its initial preparation in a Gibbs state at temperature T allows for the precise encoding of thermal information onto the probe's longitudinal relaxation via the interaction between Q1 and Q4.

The circles indicates rotation around the x-axis (blue) and y-axis (red). In addition, the strength $p$ and $q$ are controlled by the angles $\theta_1$ and $\theta_2$, respectively, according to the relations $\theta_1 = \arccos\left(1 - 2p \right)$ and $\theta_2 = \arccos\left(1 - 2q \right)$ with $\theta_1$ and $\theta_2$ ranging from $0$ to $\pi/2$. Meanwhile, the coupling between Q1 and the ancilla $Q_i$ is mediated by the interaction Hamiltonian $H_J = \left(h J/4\right) \sigma_z^{Q_1}\otimes \sigma_z^{Q_i}$ \cite{PhysRevA.106.022436}. Here, the coupling strength $J$ depends in general of the molecular structure, but we are assuming to be the same for simplicity. In addition, for the environment to be interrogated, the interrogation time $t$ is dictated by the parameter $s$ by the relation $\cos\left(s\right) = \exp\left(-\Gamma t \right/2)$. Thermal equilibrium between the probe system and the environment is achieved at $s = \pi/2$. On the other hand, for the GAD channel, the temperature is encoded in the state of Q4, and it is prepared in a thermal state such that $\rho_{Q4} = \exp\left[-\beta H_{Q4}\right]/Z_4$, with the $H_{Q4} = -\hbar\omega \sigma_z^{Q4}/2$ the qubit Hamiltonian and $Z_4 = Tr\left[e^{-\beta H_{Q4}}\right]$ the partition function.

Although we have illustrated the full quantum circuit in Fig. \ref{circuit_figure} using four qubits, the same procedure may be implemented with only two qubits, following the Refs. \cite{PhysRevA.106.022436}. In this case, a liquid sample of $^{13}\mathrm{C}$-labeled $\mathrm{CHCl}_3$ diluted in deutered
Acetone-d6 in a nuclear magnetic resonance (NRM) could be employed, with the nuclear spin of the $^{1}\mathrm{H}$ and of the $^{13}\mathrm{C}$ playing the role of the probe system and ancilla, respectively \cite{PhysRevLett.123.240601}.

\section{Conclusions and remarks} \label{Conclusions and remarks}

We investigated a state preparation framework that leverages the tunability of non-selective generalized measurements to engineer probe states capable of maximizing sensitivity in quantum sensing and thermometry. After introducing the method, we applied it to characterize parameters of relevant channels associated with thermodynamic processes. The QFI is related with different thermodynamic susceptibilities even in the transient regime. By fine-tuning the parameters of the generalized measurements, we designed probe states that achieve the best QFI for each parameters to be estimated, showing the relevant of the method. 

Specifically for the GAD channel, we briefly discussed a quantum circuit to implement the parameter estimation protocol, which could be performed in nuclear magnetic resonance. We highlight that the versatility
of the present probe preparation protocol suggests its immediate
applicability to other relevant noise models, such as the
class of Pauli channels. We hope the present work can be useful for further investigation of generalized quantum measurement in quantum parameter estimation tasks.


\begin{acknowledgments}
J. F. G. Santos acknowledges CNPq Grant No. 420549/2023-4, Fundect Grant No. 83/026.973/2024, and Universidade Federal da Grande Dourados for support. S. Su acknowledges the support by National Natural Science Foudation of China Grant No 12364008. M. Rojas would like to thank CNPq for support through Grant. No 311565/2025-5.
\end{acknowledgments}

\bibliography{ref}

\end{document}